\documentclass[
 onecolumn,12pt,
 amsmath,amssymb,
 aps,
 prfluids]{revtex4-1}
\usepackage[colorlinks=true,linkcolor=black,citecolor=blue,urlcolor=blue]{hyperref}
\usepackage{graphicx} 
\usepackage{dcolumn} 
\usepackage{ulem} 
\usepackage{bm} 

\usepackage{float} 

\newcommand{\fr}[1]{\ref{fig:#1}}
\newcommand{\er}[1]{(\ref{eq:#1})}

\begin{document}

\title{Slingshot droplet generator for clogging-free and tunable production of small uniform drops in air}

\date{\today}

\author{K. S. Surabhi}
\author{Sander Wildeman}
\affiliation{Institut Langevin, ESPCI Paris, PSL Research University, 1 rue Jussieu, 75005 Paris, France}
\keywords{droplets}

\begin{abstract}
	 Submillimeter air-borne drops of well defined size are notoriously difficult to produce in the lab. Here, we demonstrate that any simple droplet dripper, in which millimetric droplets are created by letting them drip under gravity from a hollow needle, is readily turned into a versatile droplet generator capable of producing drops smaller than the needle diameter. This is achieved by vibrating the needle base at a frequency close to a flexing resonance of the needle tube. Through high-speed imaging and detailed image analysis we show that the droplets flung off one-by-one from the needle tip are uniform in size. The produced drop size is mainly set by the driving frequency and is practically independent of needle geometry or driving amplitude, making the generator robust and easily tunable. Like for a simple dripper, the drop ejection rate is independently controlled through the flow-rate. Since the needle diameter can be much larger than the produced drop size, clogging is not an issue. By using needles with different resonant frequencies we could produce droplets with radii anywhere between 100 microns to 1 mm with a high reproducibility (the production of smaller drops is possible in principle). We propose a mass-spring model for the vibrated pendant drop that captures all our observations.
\end{abstract}

\maketitle

\section{Introduction}

Liquid drops are the drosophila of fluid dynamics research at the small scale. In many studies and applications, with topics ranging from drops walking on a vibrated bath \cite{CPFB05}, Leidenfrost drop dynamics \cite{BMBL18, GDPL19, LMWS19}, characterization of superhydrophobic surfaces \cite{TLIR13} to drop collisions \cite{QiLa97}, freezing of electrically suspended droplets \cite{LKPH18}, acoustic levitation for macromolecular crystallography \cite{MDAN19} or contactless fluid mixing \cite{WaHA18}, one wishes to create individual free (air-borne) droplets with sizes well below the capillary length (radii below 1 millimeter). However, as is well known to anyone who has experimented with droplets in the lab, it is surprisingly difficult to create such small free droplets in a consistent and automated manner.

Droplets with radii of the order of one millimeter are readily created by letting them drip, under the action of gravity, from the tip of a standard hypodermic needle. The produced drop radius $R$ in this configuration is set by a force balance between the downward pull of gravity $F_g \sim \rho g R^3$ and the surface tension force $F_{\gamma} \sim \gamma c$, by which the drop holds on to the outer circumference $c$ of the needle (where $\rho$ and $\gamma$ denote the liquid’s density and surface tension and $g$ is the gravitational acceleration). To create a ten times smaller drop by this method, a needle with a thousand times smaller outer diameter would thus be required. Besides the difficulty in fabricating such thin and fragile needles, pushing liquid out through them requires exceedingly high pressures and is prone to clogging.

Uniform submillimeter droplets are therefore most commonly produced either using drop-on-demand (DoD) type devices \cite{YCKG97, ChCh03, DoCM06, HaLB15}, in which droplets are ejected from a small hole or short nozzle by a pressure impulse, or through the (acoustically controlled) breakup of a liquid jet \cite{ScHe64, VeBJ89, LiEC90, DSDJ14}. The droplet radius in these methods is roughly set by the inner diameter of the nozzle, so that small drops are more easily produced as compared to the dripping method (requiring only a ten times smaller exit diameter for a ten times smaller drop). However, for the smallest drop sizes clogging and viscous drag is still an issue, especially when using complex non-newtonian liquids. Furthermore, both methods require specialized equipment (such as high-voltage piezo drivers) to function and often have to be operated in a continuous mode for stable output (requiring a post-selection mechanism to obtain individual drops). Various alternative drop generation techniques have been proposed over the years, including a punctured vibrating membrane \cite{PeKh03}, a liquid-air co-flow nozzle \cite{Lane47, BuMc60}, or recently, local evaporation induced pinch-off \cite{WoVL18}.

Here we propose an elegant method for producing single drops of tunable uniform size that inherits the simplicity and robustness of the gravitational dripping method. A great advantage of our method is that the produced drop size can be much smaller than the exit diameter, so that clogging and viscous drag are not much of an issue. The basic idea is shown in Fig. \fr{fig1}. Instead of letting the droplet drip from a stationary needle (Fig. \fr{fig1}A), we horizontally vibrate the needle base (Fig. \fr{fig1}B). When the frequency of this vibration is close to the flexing resonance of the needle, the tip of the needle undergoes a large amplitude motion. This causes the pendant drop to be flung off sideways at a size significantly smaller than the needle diameter (inset Fig. \fr{fig1}B). The effect would in principle be same for a stiff needle shaken at sufficient amplitude, but by exploiting the needle’s resonance a simple loudspeaker or other low cost actuator can be used. In general (depending on actuation and needle shape) the needle tip traces out an elliptical trajectory, so that the kinematics of the drop ejection resembles that of a pebble sling (Fig. \fr{fig1}C). This “slingshot” method has been proposed before for producing small water drops \cite{Dimm50, MaJW63}, but it seems that there has never been any systematic investigation into the mechanisms underlying the process or the influence of the different control parameters on the droplet size.

\begin{figure}[t]
	\includegraphics[width=.8\textwidth]{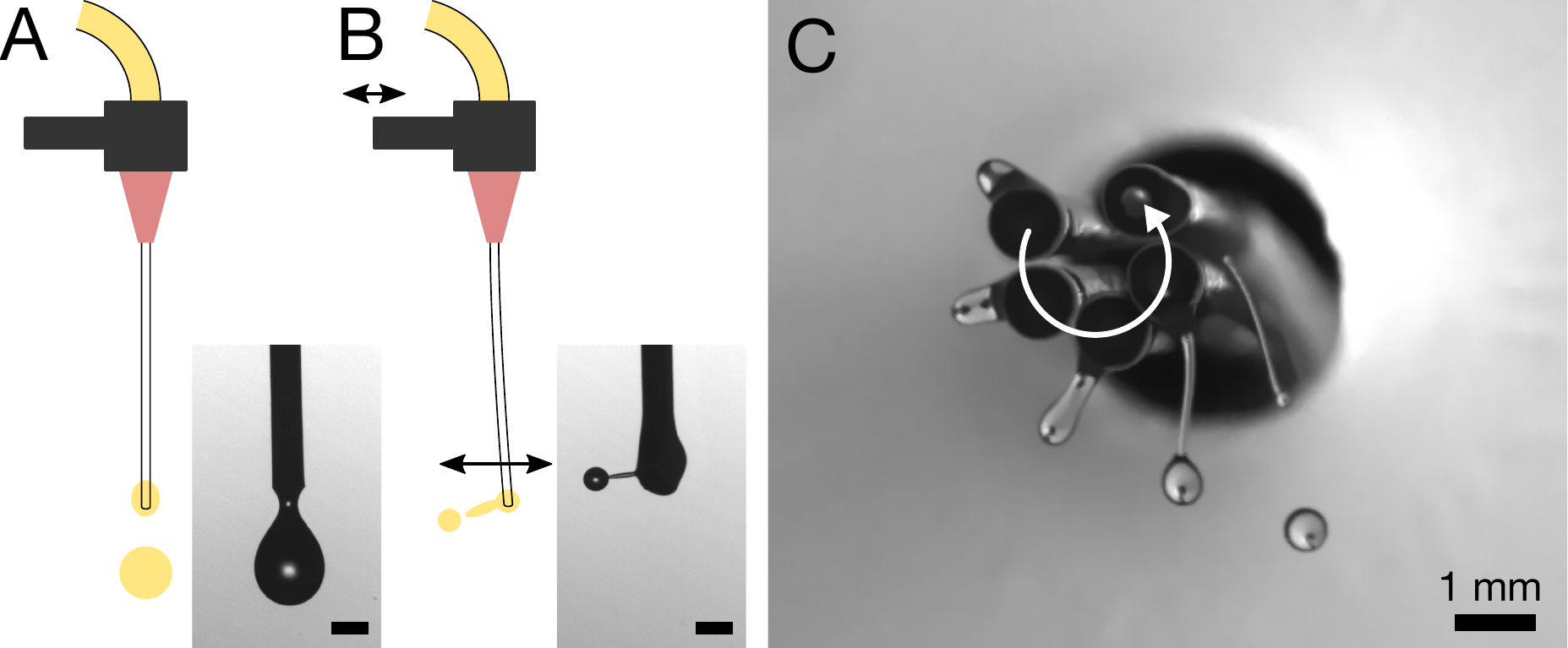}
	\caption{\label{fig:fig1} Operation principle of the slingshot droplet generator. Any conventional droplet dripper (\textbf{A}) can be turned into to a slingshot generator (\textbf{B}) by vibrating the needle base at a frequency $f$ close to the flexing resonance of the needle tube. Snapshots from the side view recordings are shown as insets (scale bar: 1 mm). \textbf{C}. Bottom view of the ejection process ($f = 87\,\text{Hz}$). Five subsequent movie frames with a time interval of 2 ms are overlapped.}
\end{figure}

\section{Experimental method}

Our experimental setup consisted of a small horizontally oriented vibration exciter (Bruel \& Kjaer type 4810) with a custom made holder into which standard hypodermic needles could be fitted vertically (Fig. \fr{fig1}B). We used both a polypropylene plastic needle (Nordson, 25 GA, outer diameter 0.85 mm) and a stainless steel needle (outer diameter 0.47 mm), which had flexural resonances (in filled condition) at 90 Hz and 600 Hz, respectively. The plastic tip had a relatively low Q factor, allowing it to be used over a wide range of frequencies (from 20 Hz to 120 Hz). The needles were fed with 20 cSt silicone oil using a small syringe pump at a constant rate of approximately 0.8 $\mu$L/s. 

The droplet detachment process was filmed under backlight illumination conditions using two high speed cameras (Photron FASTCAM Mini WX100 for side view and Basler MQ003CG-CM for bottom view). Drop size statistics were obtained from the side view at low recording rates ($\sim$1 frame per cycle) to capture at least 200 drops in a single recording. The magnification was such that 1 pixel corresponded to 14.6 $\mu$m. Shutter times were kept below 100 $\mu$s to prevent motion blur. A custom image analysis script was written to extract the droplet boundaries with sub-pixel accuracy (using gray-scale information). We checked that our boundary tracing method is robust against moderate amounts of motion and out of focus blur. The final detection resolution was estimated to be about 1 $\mu$m. An ellipse with axes perpendicular (half-width $a$) and parallel (half-width $b$) to the needle was fitted to the extracted drop boundaries. From this, the effective drop radius $R = (a^2b)^{1/3}$ was determined, assuming that the drops are slightly flattened in the direction parallel to the needle (as will be discussed below).

\section{What sets the drop size?}

In earlier works it was supposed that it is the maximum acceleration experienced by the pendant drop that mainly sets its pinch-off size [16,21]. This makes sense if one assumes that the only role of needle vibration is to increase the effective gravity experienced by the drop (in the accelerated frame of reference of the needle tip). However, we found that although a certain minimum tip acceleration $a_c$ is required to produce drops (Fig. \fr{fig2A}A), the drop size decreases only very gradually as the driving amplitude is increased beyond this value (at a fixed frequency) (see Fig. \fr{fig2A}B). Also the liquid injection rate does not seem to significantly alter the produced drop size (as long as the drop production rate stays below the needle frequency). This leaves the vibration frequency as the main candidate.

\begin{figure}
	\includegraphics[width=\textwidth]{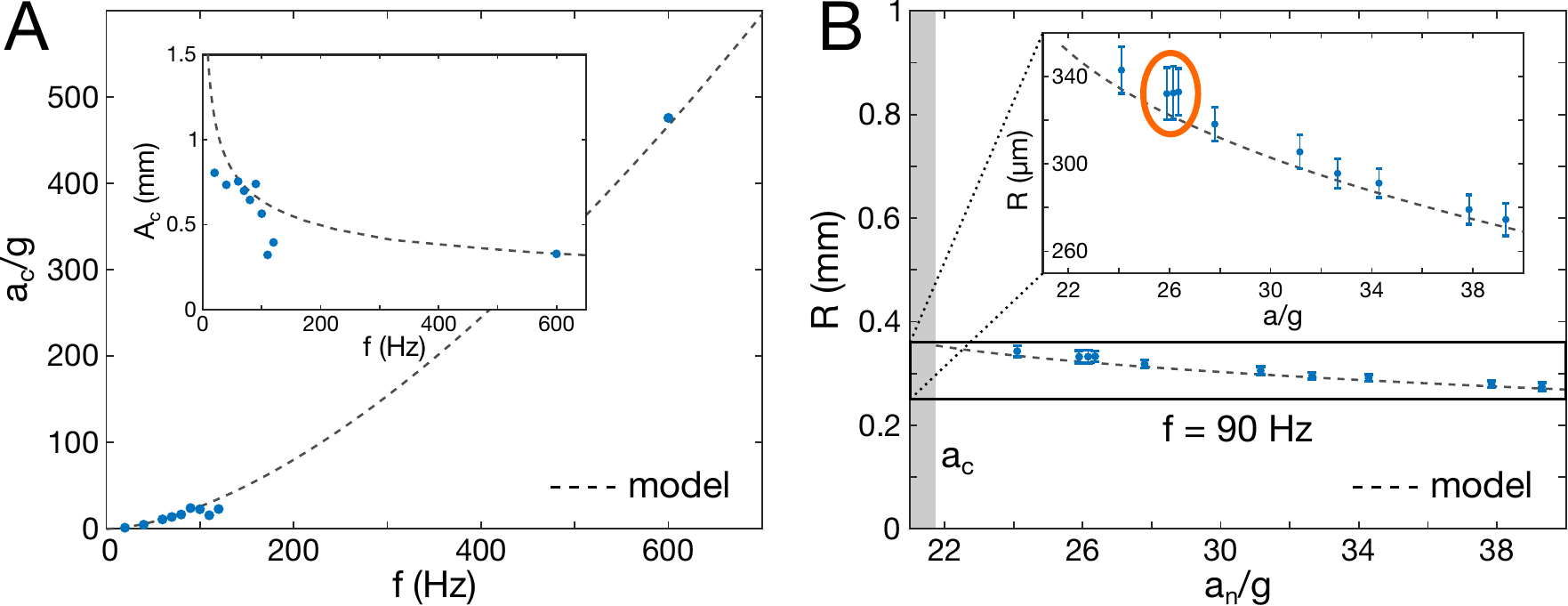}
	\caption{\label{fig:fig2A} Effect of vibration amplitude on drop production. \textbf{A.} Minimum horizontal needle tip acceleration $a_c$ required to start droplet generation as a function of actuation frequency $f$. The corresponding amplitude $A_c=a_c/\Omega^2$ is shown in the inset. \textbf{B.} Radii of produced drops as a function of the needle tip acceleration $a_n$ at a fixed frequency ($f = 90\,\text{Hz}$). Only a very slight decrease in drop radius with vibration amplitude is observed. Drop formation only occurs outside the shaded region, beyond $a_c$. Encircled data points were obtained for three different drop production rates (4, 8 and 14 drops/s).  Dashed lines show the predictions of our model.}
\end{figure}

In Fig. \fr{fig2B}A snapshots of droplet pinch-off at three different actuation frequencies (20Hz, 100Hz and 600Hz) are shown. A steep decrease in drop radius $R$ with actuation frequency $f$  can be observed. In each case the needle was driven at the critical acceleration $a_c$. As shown in Fig. \fr{fig2B}B and its inset, the $R(f)$ relation displays a clear power law behavior $R \propto f^k$ with an exponent $k = -2/3$. Notice that the data points below 120Hz and at 600Hz were obtained with needles of completely different material properties and geometry, indicating that the scaling is very robust.


The relative standard deviation $\sigma/R$ of the measured drop sizes is 2-3\% for the frequencies below 120 Hz and about 6\% for the 600 Hz case. This distribution is broader than what can be currently achieved with impulse driven DoD devices \cite{HaLB15}, but it is probably narrow enough for many applications that need a simple and robust supply of small drops (of tunable size). At a driving frequency of 110 Hz a remarkably narrow radius distribution with a relative standard deviation below 1\% was observed (see Fig. \fr{fig2B}C, central panel). At $f = 120 \,\text{Hz}$ the distribution turned out to be bi-modal (Fig. \fr{fig2B}C, bottom panel). At this frequency, 30 Hz above the needle’s resonance, two drops of somewhat different sizes were observed to be ejected in quick succession at regular intervals (see inset Fig. \fr{fig2B}C). This rich variety in drop radius statistics hints at the possibility to engineer the drop size distribution by controlling the needle actuation more precisely.

\begin{figure}
	\includegraphics[width=\textwidth]{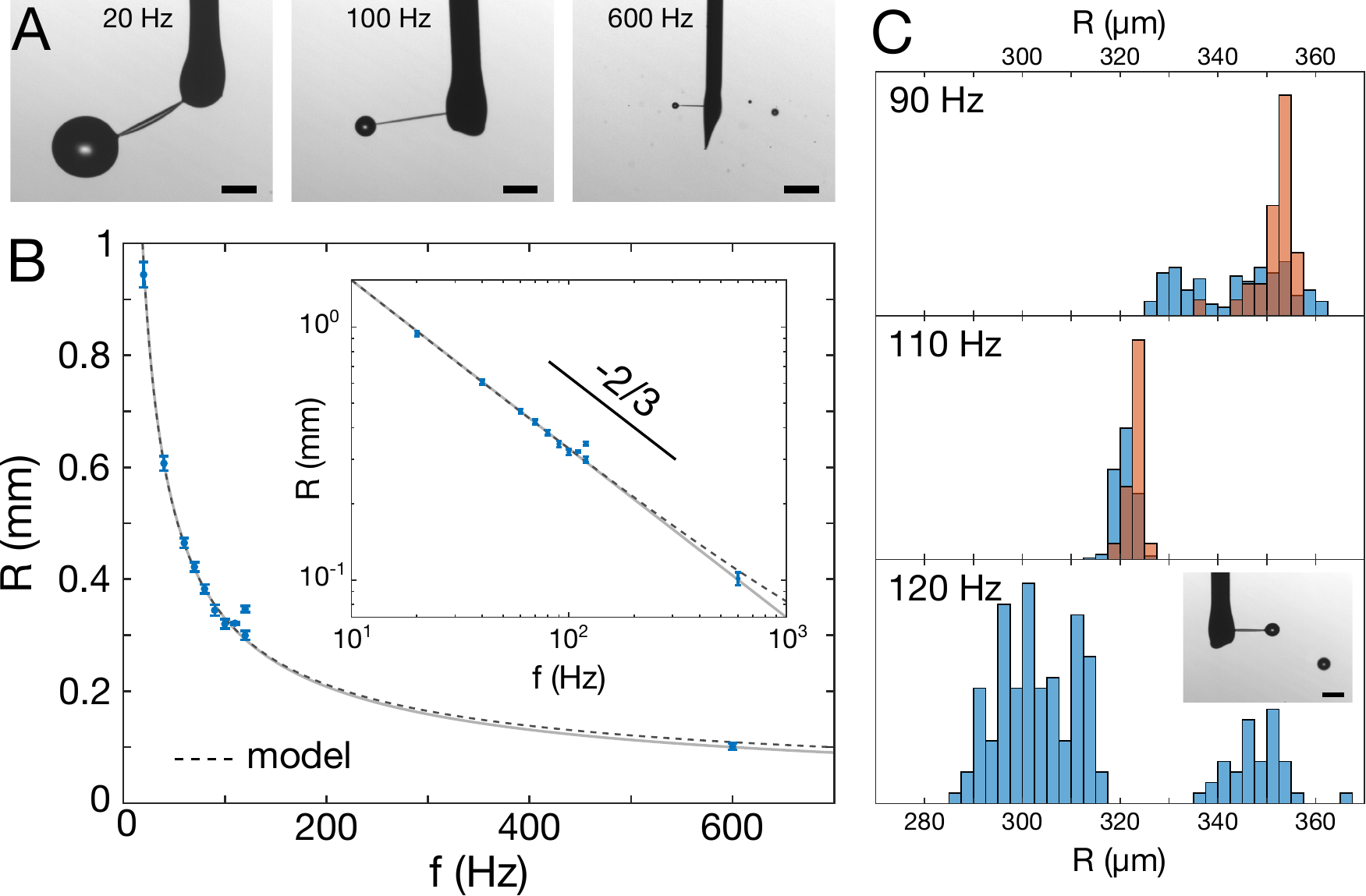}
	\caption{\label{fig:fig2B} Effect of vibration frequency on drop production. \textbf{A}. Snapshots of the droplet ejection (just before pinch-off) at three different operation frequencies (scale bars: 1 mm). \textbf{B}. Relation between drop radius $R$ and driving frequency $f$. Symbols represent experimental measurements (error bars: $\pm 1\sigma$ of drop distribution). The solid line is a fit to Eq. \er{scaling} and the dashed line represents the extended model including viscosity. \textbf{C.}  Measured drop size distributions for frequencies at and above needle resonance. Blue histogram bars represent the distribution of all measured drops. For the red histograms only drops flying off in a plane approximately parallel to the imaging plane were selected.}
\end{figure}

We also investigated the effect of selecting drops based on the direction in which they fly off the needle. The red histogram bars in Fig. \fr{fig2B}C show the distribution of drops that show minimal out-of-focus blurring as they move away from the needle (corresponding to drops that fly off in a direction parallel to the imaging plane). The distribution of these drops is significantly more narrow than that of the unfiltered drops, indicating that part of the distribution width can be attributed to slightly different detachment conditions in different directions.

\section{Pinch-off condition}

We can explain our observations with a simple mass-spring type model taking into account the inertia-capillary dynamics of the vibrated pendant drop (Fig. \fr{fig3}A). Similar to a freely oscillating droplet \cite{Stru96} the small fluid mass collected at the end of the tip will have a natural oscillation frequency $\omega_0$  scaling as $\omega_0^2 \sim \gamma / (\rho R^3)$. As the attached drop volume increases with time the drop’s natural frequency will eventually approach the needle’s angular frequency $\Omega = 2 \pi f$. Each time the resonant condition $\omega_0 \approx \Omega$ is reached the oscillation amplitude of the drop will diverge, causing it to detach. This model thus predicts the radius of the produced drops to be given by
\begin{equation}\label{eq:scaling}
R^3 = c \frac{\gamma}{\rho \Omega^2}\text{,}
\end{equation}
where $c$ is a pre-factor of $\mathcal{O}(1)$ which could depend weakly on the needle geometry, driving amplitude and viscosity of the liquid. Setting $\gamma = 20.6\,\text{mN/m}$ and $\rho = 950 \,\text{kg/m$^3$}$ (taken from the datasheet of the used silicon oil) we find that Eq. \er{scaling} fits the data very well with a prefactor of $c \approx 0.65$ (see solid line in Fig. \fr{fig2B}B). Interestingly, Eq. \er{scaling} resembles that for the median droplet size produced in ultrasonic nebulizers \cite{KACB19}, except that $\Omega$ has to be replaced by $\Omega/2$ in the case of the nebulizer, as the forcing is parametric in that case.

\section{Mass-spring dynamics of the pendant drop}

\subsection{Experimental validation}

To demonstrate the mass-spring behavior of the pendant drop in a more direct manner we tracked the horizontal locations of the drop and the needle as the drop slowly accumulates mass (Fig. \fr{fig3}B). As shown in Fig. \fr{fig3}C the droplet position initially closely follows the motion of the needle, but once it reaches a certain critical mass its oscillation amplitude rapidly increases. This sudden divergence of the amplitude is more apparent in Fig. \fr{fig3}D, where we show the oscillation amplitude of the drop in the frame of reference of the needle tip. For a harmonic oscillator driven at resonance this amplitude divergence should be accompanied by a phase lag approaching 90$^\circ$ with respect to the forcing. Such an increase in phase lag can indeed be observed (Fig. \fr{fig3}C), providing strong support for the mass-spring model. Careful inspection of the traces in Fig. \fr{fig3}B reveals that the oscillation amplitude of the needle slightly reduces just before the drop detaches, evidencing the efficient energy transfer from the needle to the drop at this point.

\begin{figure}
	\includegraphics[width=.9\textwidth]{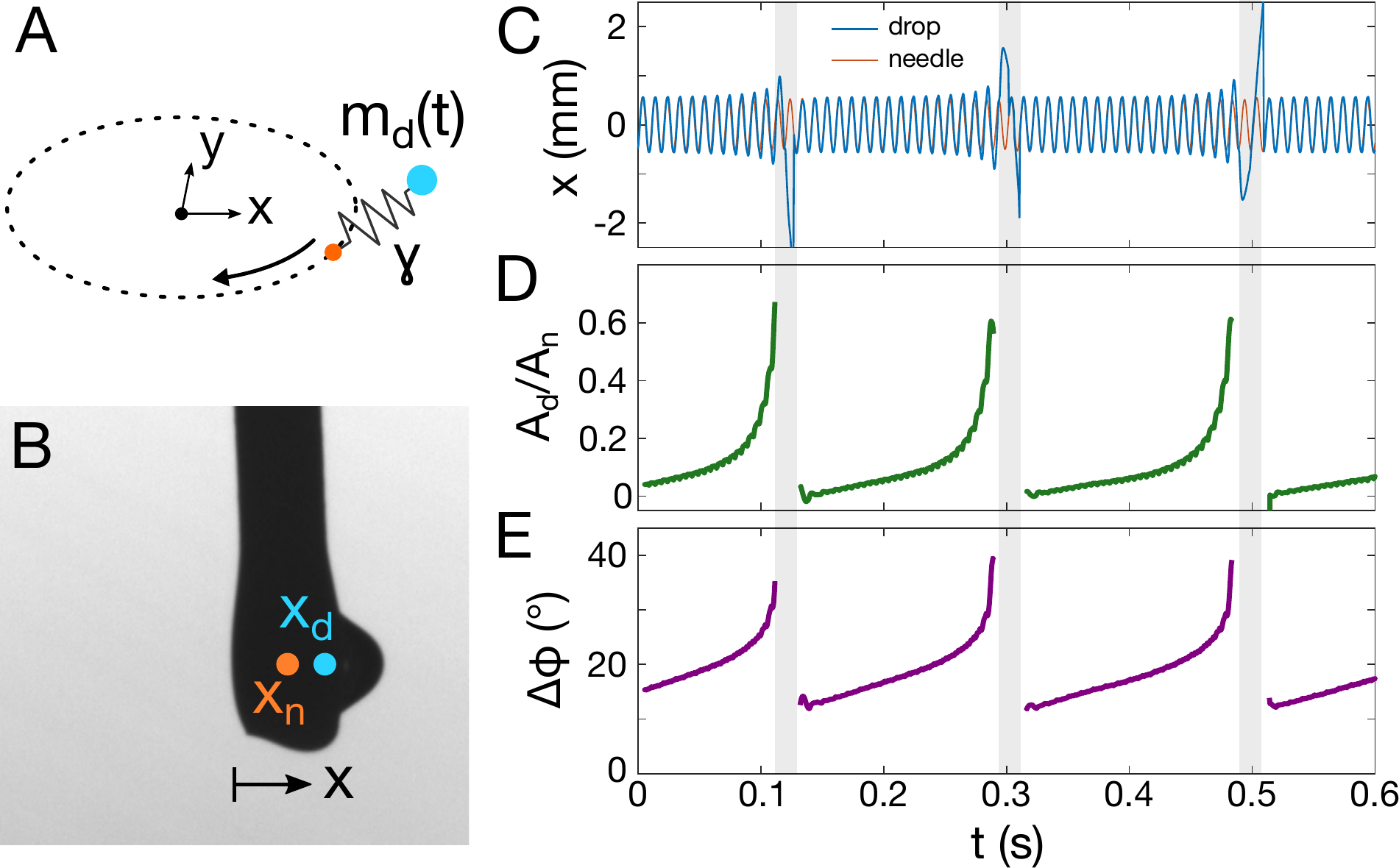}
	\caption{\label{fig:fig3} Mass-spring model for the pendant drop. \textbf{A.} Schematic of the model. A slowly increasing fluid mass (blue dot) is attached to the oscillating needle tip (orange dot) by a surface tension spring. \textbf{B.} Experimental definition of the needle tip position $x_n$ and pendant drop position $x_d$ in the side view recordings. \textbf{C.} Time trace of the extracted drop and needle positions in the lab frame. The gray shaded regions mark pinch-off events. \textbf{D.} Oscillation amplitude of the drop (normalized by the needle amplitude) in the frame of reference of the needle tip. \textbf{E.} Phase lag of the drop oscillation with respect to the needle oscillation.}
\end{figure}

\subsection{Role of viscosity}

So far we have neglected the role of viscosity in the drop generation process. Viscosity can be incorporated in the mass-spring picture as a damping term $F_d \sim \mu R v$, where $v$ is the velocity of the liquid mass relative to the needle tip and $\mu$ is the dynamic viscosity of the liquid. With this, an approximate equation of motion for the displacement $\delta x$ of the pendant drop (in the accelerated frame of reference of the needle tip) can be written as
\begin{equation}\label{eq:model}
\rho R^3 \ddot {\delta x} + \beta \mu R \, \dot{\delta x} + \alpha \gamma \, {\delta x} = \rho R^3 \Omega^2A_n \cos(\Omega t)\text{,}
\end{equation}
where $\alpha$ and $\beta$ are geometric prefactors of $\mathcal{O}(1)$ and $A_n$ is the oscillation amplitude of the needle tip. By introducing $R_0 = (\alpha \gamma / (\rho \Omega^2))^{1/3}$ (the resonant radius for $\mu \rightarrow 0$) as length scale and $\Omega^{-1}$ as time scale, this equation is cast in the dimensionless form
\begin{equation}\label{eq:moddl}
\tilde R^3 \ddot X + \frac{\beta}{\sqrt{a}}\text{Oh} \tilde R\dot X + X = \tilde R^3 \tilde A_n \cos(T)\text{,}
\end{equation}
wherein $X = \delta x/R_0$, $\tilde R = R/R_0$, $\tilde A_n = A_n/R_0$, $T = \Omega t$ and the Ohnesorge number $\text{Oh} = \mu / \sqrt{\rho \gamma R_0}$ appears as a parameter. In our experiments Oh ranges from 0.1 for the largest drops to 0.4 for the smallest drops. 

The steady state solution to Eq. \er{moddl} is $X = \tilde A_d \cos(T + \phi)$, where $\phi$ is a phase lag with respect to the driving and $\tilde A_d = A_d/R_0$, the dimensionless drop oscillation amplitude, is given by the resonance curve
\begin{equation}\label{eq:modsol}
\tilde A_d = \frac{\tilde R^3 \tilde A_n}{\sqrt{\left(1-\tilde R^3\right)^2+(\beta^2/\alpha)\text{Oh}^2 \tilde R^2}}\text{.}
\end{equation}

If we assume that pinch-off is triggered when $\tilde A_d(\tilde R) \sim \tilde R$, then Eq. \er{modsol} predicts both the existence of a threshold amplitude for drop production and a decrease in drop radius for a further increase in driving strength. For $\text{Oh} < 1$, the elongation $E = \tilde A_d/\tilde R$ reaches a maximum value of $E^* \approx (\sqrt{\alpha}/\beta) \text{Oh}^{-1}\tilde A_n (1 + \epsilon/2)$ at a radius of $\tilde R^* \approx (1 + \epsilon)$, where $\epsilon = \beta^2 \text{Oh}^2 / (9 \alpha)$ represents a small shift in resonance condition as compared to the non-viscous case. By requiring $E^* \sim 1$, we obtain that the threshold vibration amplitude $A_c$ should vary as $A_c \sim \text{Oh} R_0 \sim \mu (\gamma \rho^2 \Omega)^{-1/3}$, or in terms of the needle acceleration $a_c = \Omega^2 A_c \sim \mu \Omega^{5/3} (\gamma \rho^2)^{-1/3}$. We have found that with $\alpha = 0.75$ and $\beta = 3.3$ and by taking $E > 2$ as a pinch-off criterium (an elongation greater than one drop diameter) the model simultaneously captures the observed dependency of drop size and critical vibration amplitude on frequency (Figs. \fr{fig2A}A and \fr{fig2B}A) and the gradual variation of the drop size with vibration amplitude (Fig. \fr{fig2A}B). For a good quantitative agreement we had to assume that stable drop ejection only occurs when the driving strength is about 10\% above the absolute minimum.

\section{Drop rotation after pinch-off}

\begin{figure}
	\includegraphics[width=.9\textwidth]{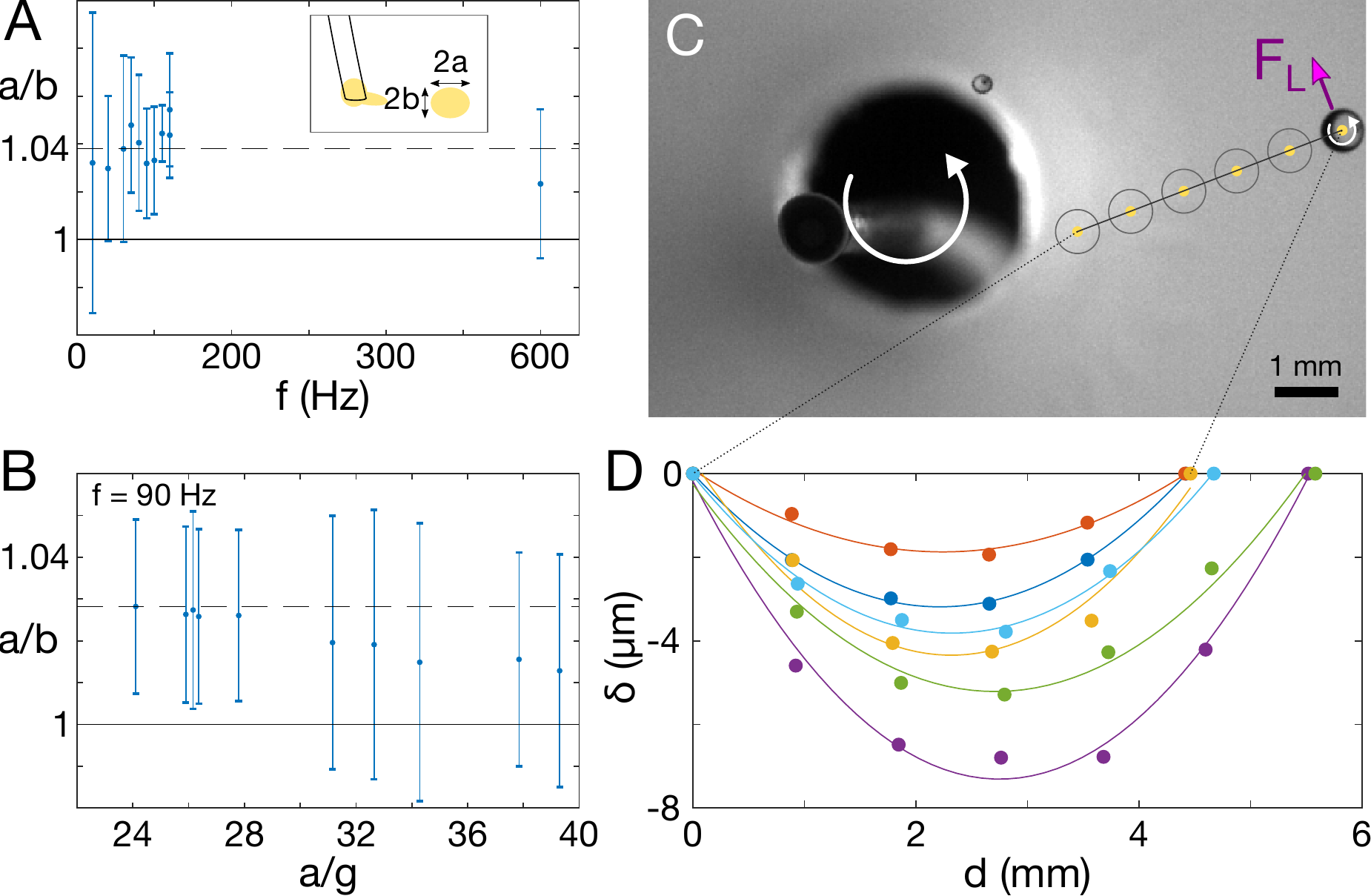}
	\caption{\label{fig:fig4} Evidence of drop rotation after pinch-off. \textbf{A.} Drop aspect ratio as a function of forcing frequency. The dashed line indicates the mean value. \textbf{B.} Drop aspect ratio for increasing needle acceleration at $f =  90\,\text{Hz}$. (dashed line is a guide to the eye). \textbf{C.} Example of transverse drop trajectory extracted from the bottom view. \textbf{D.} Detailed view of 6 measured trajectories, rotated such that the x-axis coincides with the first and last point on the trajectory. The yellow data points correspond to the trajectory in C. Parabolas were fitted to the trajectories to extract their radius of curvature. The spin and lift force directions consistent with the observed deflection are indicated in C.}
\end{figure}

Finally, we have investigated the state of motion of the drops after they have left the needle. As mentioned above, the tip of the needle generally describes an elliptical trajectory. An interesting byproduct of this circular motion is that it can impart some angular momentum to the detaching drop. Detailed analysis of the droplet contours while the drops are in flight reveals that on average the drop shape is not exactly spherical (Fig. \fr{fig4}A). They bulge out slightly in the plane perpendicular to the needle with an ellipticity $\kappa$ of about 4\%. This shape is consistent with a drop spinning at roughly the angular frequency $\Omega$ of the needle. A balance between the centrifugal force in a droplet spinning with frequency $\Omega_d$ and surface tension leads to $\kappa = (1/8)\rho R^3 \Omega_d^2/\gamma$ \cite{Rayl14}. Combining this with Eq. \er{scaling} for the size of the drops we obtain $\kappa = (c/8)\Omega_d^2/\Omega^2 \approx 0.08\, \Omega_d^2/\Omega^2$. Thus the observed ellipticity of 4\% can be explained by a drop spin rate of $\Omega_d \approx 0.7 \,\Omega$. A slight decrease in ellipticity with an increase in forcing amplitude was observed (Fig. \fr{fig4}B).

In addition to this flattening, a spinning droplet moving with velocity $v$ through air (density $\rho_a$) should experience a lift force $F_L \approx \pi \rho_a R^3 \Omega_d v$ (valid for small Reynolds numbers) perpendicular to its trajectory \cite{RuKe61}. Balancing this with the required centripetal force $(4\pi/3)\rho R^3 v^2/r$ for an orbit with a radius $r$, implies that the rotating drop should follow a curved trajectory with $r \approx (4/3)(\rho/\rho_a) v /\Omega_d \approx (\rho/\rho_a) v /\Omega$. Figure \fr{fig4}C shows a bottom view with a droplet ejected at a velocity of about $v = 0.45\,\text{m/s}$ for a needle frequency of $\Omega = 565\,\text{rad/s}$. Taking the air density to be $\rho_a = 1.2\,\text{kg/m$^3$}$, we predict $r \approx 0.6\,\text{m}$. Although such a gradual deflection is difficult to observe by eye, a detailed image analysis (Fig. \fr{fig4}D) indeed reveals a slight curvature $r_\text{exp} \approx 0.75 \, \text{m}\pm 0.25\,\text{m}$ (averaged over 6 consecutive trajectories) consistent with this prediction.

We note that the observed small drop flattening and transverse deflection are only just within our detection limits. To ascertain that the droplet rotates, a more direct observation (e.g. by inserting tracer particles) would be favorable. This is beyond the scope of the current study however.

\section{Conclusion and outlook}

To conclude, the slingshot droplet generator produces drops of a nearly uniform size set by the condition that their Rayleigh oscillation frequency matches the needle actuation frequency. A minimum oscillation amplitude is required to overcome the viscous dissipation in the drop, but otherwise the size depends only very weakly on the amplitude. We have found evidence that some angular momentum can be imparted to the free drops.

The drop size can vary somewhat with the ejection direction. We suspect that this anisotropy can be either amplified or suppressed by controlling the needle trajectory in more detail (e.g. by controlling the ellipticity of the needle trajectory with two perpendicular actuators at the needle base). Other modes of operation (such as the bimodal ejection observed at frequencies above the needle’s resonance) could perhaps be found by exploring the effects of flow rate and frequency detuning over a wider parameter range. 

An appealing aspect of the production process is that the main droplet and the accompanying satellite drop are spatially separated. They fly off in different directions and the smaller drops have a smaller horizontal travel due to air drag. This makes it possible to tap into one stream or the other by simple spatial filtering \cite{Dimm50, MaJW63}. The unique sideways droplet ejection can be advantageous in applications that need a gentle deposition onto a (liquid) surface or into an acoustic or electric trap. 

An array of generators could be used to produce clouds of drops with a highly tunable size distribution, which may be beneficial in the challenging experimental study of natural cloud dynamics \cite{VMWP99}. 

Finally, the high needle diameter to drop size ratio that can be attained, could make the slingshot method especially suitable for the production of small drops of complex composition.


%

\end{document}